\documentstyle[twocolumn,prl,psfig,aps]{revtex}
\begin{document}
\twocolumn[\hsize\textwidth\columnwidth\hsize\csname @twocolumnfalse\endcsname 
\draft
\begin{title}
{Even-odd behavior of conductance in monatomic sodium wires}
\end{title}
\author{H.-S. Sim,$^{1,2}$ H.-W. Lee,$^3$ and K. J. Chang$^{1,3}$}
\address{$^1$Department of Physics, Korea Advanced Institute of
Science and Technology, Taejon 305-701, Korea}
\address{$^2$Max-Planck-Institute for the Physics of Complex Systems, 
N\"{o}thnitzer Str. 38, D-01187 Dresden, Germany}
\address{$^3$School of Physics, Korea Institute for Advanced Study,
207-43 Cheongryangri-dong, Dongdaemun-gu, Seoul 130-012, Korea}
\date{\today}
\maketitle
\begin{abstract}
\widetext
With the aid of the Friedel sum rule, we perform first-principles
calculations of conductances through monatomic Na wires, taking into
account the sharp tip geometry and discrete atomic structure of
electrodes. 
We find that conductances ($G$) depend on the number ($L$)
of atoms in the wires; $G$ is $G_0 (= 2e^2/h)$ for odd $L$, independent
of the wire geometry, while $G$ is generally smaller 
than $G_0$ and sensitive to the wire structure for even $L$.
This even-odd behavior is attributed to the charge neutrality and 
resonant character due to the sharp tip structure.
We suggest that similar even-odd behavior may appear in other monovalent
atomic wires.
\end{abstract}

]
\narrowtext

Atomic contacts have been generated by scanning tunneling microscopes
or mechanically controllable break-junctions \cite{Ruitenbeek99cond-mat}.
It has been demonstrated that single atom
contacts made of monovalent atoms such as Na, K, and Au have a strong
tendency towards the quantized conductance $G_0$ ($=2e^2/h$).
To explain the quantization, a ballistic transport model has been
suggested \cite{Olesen}, where atomic contacts are modeled by jellium
constrictions, which are adiabatically connected to electrodes,
analogous to quantum point contacts (QPC)
in two-dimensional electron gas systems \cite{Wees,Glazman}. 
However, this analogy may be inappropriate for sharp tip structures
of atomic contacts \cite{Krans95PRL}, because the Fermi wavelengths
in metallic systems are comparable to atomic spacings.
Recently, Yeyati and his coworkers \cite{Yeyati} have suggested
a resonant transport as an alternative explanation, based on
their tight-binding calculations assuming the local charge
neutrality.
Despite several theoretical attempts
\cite{Yeyati,Emberly,Brandbyge,Kobayashi}, 
the origin of the quantized conductance has not been clearly understood 
because the validity of simplifications such as
the local charge neutrality assumption \cite{Yeyati,Brandbyge} or 
a jellium electrode model \cite{Kobayashi}, 
which ignores the sharp tip geometry for electrodes, 
is rather unclear.

In this Letter, to clarify the origin of the conductance quantization, 
we calculate conductances through monatomic Na wires \cite{Nakamura},
taking into account the sharp tip shape and discrete atomic
structure of electrodes.
We perform real-space multigrid electronic structure calculations within
the local-density-functional approximation (LDA).
Using the Friedel sum rule \cite{Datta,Lee}, which relates the conductance
to the density of states (DOS), we are able to reduce greatly computational
demands.
We find a resonant character in transport and a robust quantization of
conductances when the number ($L$) of atoms in wires is odd.
For even $L$, conductances are {\it not} quantized, resulting in even-odd
behavior in transport.
This feature is very different from the length-independent
conductances observed in ballistic quantum wires.

\begin{figure}
\centerline{\psfig{figure=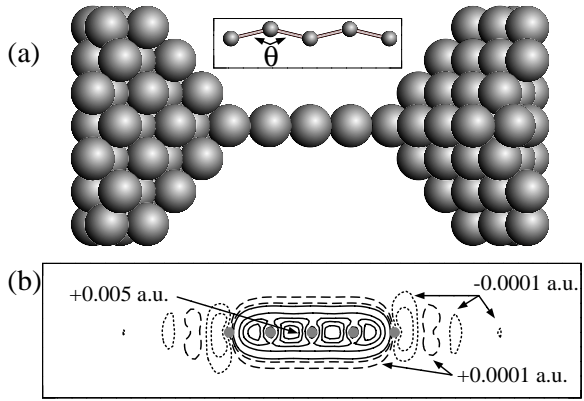,
height=,width=0.45\textwidth}}
\caption{
(a) The atomic structure for the $L=5$ Na wire connected to electrodes,
with the inversion symmetry. Each electrode is modeled by a cluster of
$M$ (=64) Na atoms.
(b) Contour plot of the difference of the total charge densities
between the ($L=5,M=64$) wire and its reference system: solid contours
for 0.001 to 0.005 a.u. with the increment of 0.001 a.u., whereas dotted
and dashed contours for -0.0003 to 0.0003 a.u. with the increment of
0.0002 a.u. Atomic positions in the wire are marked by gray dots.
A zigzag wire is shown in the inset in (a).  }
\label{fig1}
\end{figure}

Figure \ref{fig1}(a) shows a monatomic Na wire connected to two electrodes, 
which have an inversion symmetry.
Each electrode is modeled by a cluster of $M$ Na atoms in the bcc structure.
The (111) direction of the bcc lattice is aligned with the wire with $L$
Na atoms, which consists of 2 apex atoms in the clusters and $L-2$ Na 
atoms between the clusters.
For the interatomic spacing $d$ within the wire, we use the bond distance
of $d_0 = 3.659$ {\AA} in bcc bulk Na, and also test various different 
values for $d$.
Using the inversion symmetry, the DOS can be decomposed into even ($\rho_e$)
and odd parity ($\rho_o$) components.
When only one eigenchannel contributes 
to transport \cite{vandenBrom,Kobayashi},
conductances through the wire can be expressed in terms of $\rho_e$ and
$\rho_o$ using the Friedel sum rule \cite{Datta,Lee};
\begin{eqnarray}
G={2e^2 \over h} \sin^2 \left[
{\pi \over 2} (N_e-N_o) \right],
\label{conduc} 
\end{eqnarray}
where $N_{e(o)}$ [$= \int^{E_F} dE\, \rho_{e(o)}(E)$] denotes 
the number of electrons with even (odd) parity 
and $E_F$ is the Fermi energy.
With Eq.~(\ref{conduc}), we can reduce greatly the computational demand
and analyze conductance characteristics in terms of the DOSs. 

We calculate self-consistently the electronic energies of the wire system
using the real-space multigrid method \cite{Jin} within the LDA.
Norm-conserving pseudopotentials are generated by the scheme of Troullier
and Martins, 
and then transformed into the separable form of Kleinman 
and Bylander \cite{Troullier_Kleinman}.
We use a supercell geometry containing two electrodes, each of which has
a cross section of 20.76 {\AA} wide.
Electrodes in neighboring supercells are separated 
by more than 11.64 {\AA}, so that 
inter-supercell interactions are negligible.
Using a grid spacing of 0.43 {\AA}, we ensure that the total energy is
converged to within 10$^{-7}$ Ry.
To calculate the DOS, we use the Fermi-Dirac broadening of $E_T$ = 0.052 eV
(600 K) for the occupation of each level.
We find that the level splitting between even and odd parity
states, which have negligible wire characteristics, is less than 0.003 eV.
Since this value is much smaller than $E_T$ and the average level spacing
near $E_F$, we do not expect numerical errors in $N_e-N_o$ due to the
charge sloshing problem~\cite{Payne}. 

\begin{figure}
\centerline{\psfig{figure=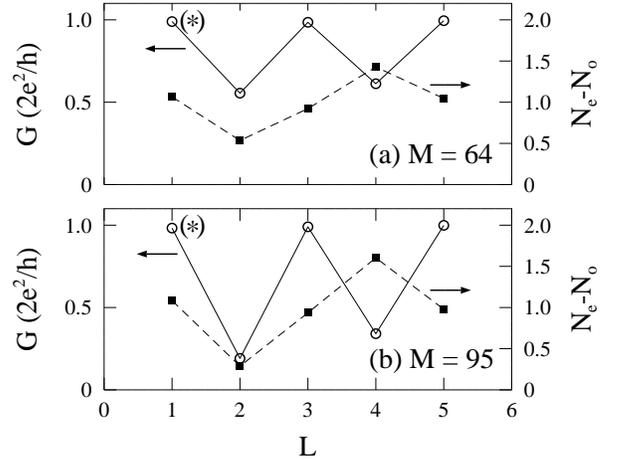,
height=,width=0.45\textwidth}}
\caption{
For unstretched linear wires with (a) $M=64$ and (b) 95, $G(L)$
(open circles) and $N_e-N_o$ (filled squares) are plotted as a function
of $L$. 
For $G(1)$ marked by asterisks, its value may be modified due to
the possibility of multi-channel transport (see text). }
\label{fig2}
\end{figure}

The conductances calculated from the DOSs are plotted as a function of $L$
for $M=64$ and 95 in Fig. \ref{fig2}. In both cases $G$ oscillates
with $L$, exhibiting even-odd behavior.
For odd $L$, we obtain $G = G_0$ 
since the difference ($N_e-N_o$) almost equals 1.
For both $M=64$ and 95, similar results are found, 
indicating that the finite size effect of clusters is negligible for odd $L$.
For even $L$, on the other hand, $N_e-N_o$ is not equal to 1 and
conductances deviate considerably from $G_0$.
Since the results for $M=64$ and 95 are different from each other,
the finite size effect seems to be more important for even $L$.
At this point, it is beyond our computational capability to 
calculate conductances for larger cluster sizes.
However, based on the DOS analysis given below, 
we confirm that the qualitative feature of $G$ being
smaller than $G_0$ for even $L$ is not an artifact
due to the finite size effect.

\begin{figure}
\centerline{\psfig{figure=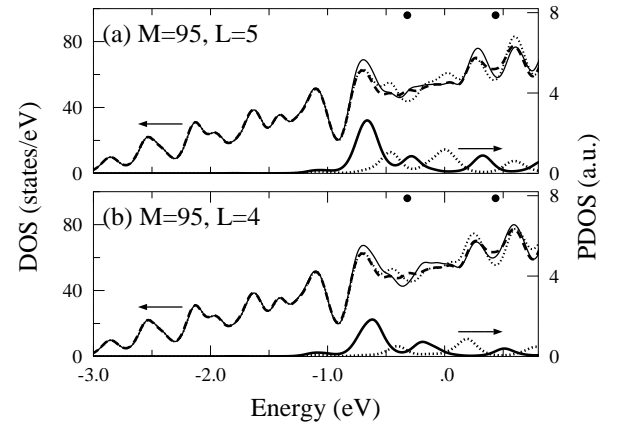,
height=,width=0.45\textwidth}}
\caption{
For a linear wire with $d=d_0$, $\rho_e(E)$ (solid lines) and
$\rho_o(E)$ (dotted lines) are compared with
$\rho^{\rm R}_e(E)[=\rho^{\rm R}_o(E)]$ (dashed lines).
The Fermi level ($E_F$) is aligned at 0 eV.
Black circles denote the locations of the resonance states which are
localized at the apex atoms of the reference system.
Thick solid and dotted lines denote the projected DOSs (PDOSs) with 
even and odd parities, respectively. 
}
\label{fig3}
\end{figure}

The DOS analysis provides useful information. From 95-atom 
cluster calculations, the results for $\rho_e(E)$ and $\rho_o(E)$ 
are drawn for the $L=4$ and 5 wires in Fig. \ref{fig3}, and compared with
$\rho_e^{\rm R}(E)$ and $\rho_o^{\rm R}(E)$ for their reference systems 
(hereafter denoted by the superscript R), which are defined by removing 
$L-2$ central atoms in the wire without altering the inter-cluster distance. 
We find that $\rho_{e(o)}$ and $\rho_{e(o)}^{\rm R}$ are almost identical 
over a wide range of energies 
except for near the Fermi level. From the projected DOSs (PDOS) 
onto the $L-2$ central atoms in the wire
(here a sphere with the radius of 0.5 $d$ is chosen for each atom),
the difference between $\rho_{e(o)}(E)$ and $\rho_{e(o)}^{\rm R}(E)$
is found to be mainly caused by the existence of resonance states.
Thus, unlike conventional QPCs, the monatomic wires have the resonant
character, consistent with previous results \cite{Yeyati}.

We estimate the spatial extension of the resonance states 
by comparing the charge densities of the wire system ($L=5, M=64$)
with those for its reference system [see Fig. \ref{fig1}(b)].
Besides the major difference in the wire region, 
an oscillatory feature due to the charge rearrangement appears on
a few layers in each electrode, in good agreement with the previous
calculations \cite{Sablikov}.
Thus, the resonance states are not strictly localized in the wire but
extended somewhat into the electrodes.

In the PDOS, there exists a half-filled resonance state 
at $E_F$ for odd $L$ (on-resonance), while the Fermi level lies between
two resonance states for even $L$ (off-resonance).
The position of the resonance states affects $N_e-N_o$, and thus
conductances.
For demonstration, we use the relation 
$N_e-N_o=(N_e-N_e^{\rm R})-(N_o-N_o^{\rm R})$, 
where $N_e^{\rm R}=N_o^{\rm R}$ because $G^{\rm R}$ = 0
in the reference system.
Since each resonance contains exactly one extra energy level \cite{Harrison},
each filled resonance state with even (odd) parity increases 
$N_{e(o)}-N_{e(o)}^{\rm R}$ by 2 without altering $N_{o(e)}-N_{o(e)}^{\rm R}$,
and thus increases (decreases) $N_e-N_o$ by 2; 
the factor two denotes spin degeneracy.
However, the filled resonance states are irrelevant because
$G$ does not change when $N_e-N_o$ changes by $\pm 2$, as shown in
Eq. (\ref{conduc}).
On the other hand, the half-filled resonance state changes 
$N_e-N_o$ by $\pm 1$, and if there is such a state, $G$ equals $G_0$. 
When the Fermi level is located at the tails of resonances,
$|N_e-N_o|$ can be quite different from $1$ and $G<G_0$.
This explains the correlation between
the even-odd behavior of conductances and
the relative position of resonance states with respect to $E_F$.

A recent tight-binding study \cite{Yeyati} for $L=1$ 
has demonstrated that the position of resonance states
cannot be arbitrary when the {\it local} charge neutrality is assumed.
Although this assumption is violated in atomic scale
due to the charge oscillation in Fig.~\ref{fig1}(b),
we find that the charge neutrality does hold globally near the wire
and constrains the resonance position for general $L$.
Since Na is a monovalent atom, the wire system contains $\Delta N$ ($= L-2$)
extra electrons, as compared to its reference system.
On the other hand, we verify that the numbers of filled resonances
in the two systems differ by $\Delta N/2$ \cite{neutrality}, which
is a half-integer for odd $L$.
This matching indicates that all extra electrons are 
in the resonance states near the wire.
Thus, the alternation of on- and off-resonances with $L$ results from
the charge neutrality.

The same even-odd behavior is expected for a wide class of wire structures,
since the resonant character and the charge neutrality are common features.
We examine various wire structures such as stretched wires with 
$d$=1.1 and 1.2 $d_0$ and zigzag wires with the bond angles
$\theta=150^\circ$ and $120^\circ$ ($d=d_0$) [see the inset in Fig.~1(a)].
In all cases, a half-filled resonance state is found for odd $L$
and $G=G_0$ to within 2 \%,
while the off-resonance transport is realized for even $L$ and $G<G_0$.
Thus, the same even-odd behavior holds for other wire structures.
On a quantitative level, however, $G$ is sensitive to the wire structure
for even $L$.
For the ($L=4,M=64$) system, for example, $G/G_0$ is 0.61 for a linear
unstretched wire, 0.50 for a stretched wire with $d=1.2 d_0$,
and 0.71 for a zigzag wire with $\theta=120^\circ$.
Recalling that in the off-resonance transport,
$G$ depends not only on the relative position of
resonance states but also on the ratio $\Gamma/\Delta E$,
where $\Delta E$ and $\Gamma$ denote 
the resonance spacing and width, respectively,
nonuniversal $G$ values for even $L$ can be understood
since the ratio depends on wire structure.
For sufficiently large $L$, however,
$G$ becomes universal for even $L$ as well
since $\Delta E\rightarrow 0$ and $G\rightarrow G_0$
as $L\rightarrow \infty$~\cite{Lee01PRB}.
In this limit, the even-odd behavior disappears.

Here we note that two key ingredients responsible
for the even-odd behavior are the sharp tip structure
and charge neutrality, which are believed to
be common in other monovalent metallic wires.
It is then very plausible that 
the same even-odd behavior may appear 
in other monovalent metallic wires.
Previous experimental studies~\cite{Ruitenbeek99cond-mat,Scheer}
showed that atomic contacts made of various monovalent metals
have many properties in common.
In agreement with our expectation, recent calculations \cite{Hakkinen}
for gold wires found that $G$ is smaller than $G_0$ and sensitive to
wire structure for $L=4$.
Experimentally, gold wires with $L>1$ have been already reported
\cite{Yanson,Ohnishi}.
Although these experiments did not exhibit an evidence for the
even-odd behavior, it is premature to reject the possibility 
of the even-odd behavior in gold wires because it is not clear
whether uncontaminated gold wires with even and odd $L$ are
both produced in the experiments \cite{gold_wire_comment}.
More experimental studies are required to clarify the existence of the
even-odd behavior.

We next discuss briefly atomic wires made of multivalent atoms. 
Since multiple eigenchannels contribute to
transport, Eq.~(\ref{conduc}) should be
replaced by the generalized formula~\cite{Datta},
$G=G_0$$\sum_{j=1}^{J}$$\sin^2[(\pi/2)(N_e-N_o)\alpha_j]$,
where $J$ is the number of channels and $\sum_{j=1}^{J}\alpha_j=1$. 
One crucial difference from monovalent wires is
that the new parameter $\alpha_j$ does depend on wire structure.
Thus, even when the charge neutrality fixes $N_e-N_o$,
$G$ still depends on wire structure.
This explains why multivalent metallic wires such as Al
\cite{Ruitenbeek99cond-mat} do not exhibit clearly quantized conductances
as in monovalent wires.
We point out that for $L$ = 1, the multichannel transport may be relevant
even for sodium wires.
We find that the coupling between two electrodes is not negligible
even when the shared apex atom is removed. 
This implies that the direct transport between the electrodes,
which is not mediated by the common apex atom, is not negligible,
in agreement with Ref.~\cite{Kobayashi}.
For more precise conductance calculations for $L=1$,
the generalized formula should be used, which is beyond
the scope of our paper.
In recent experiments for Na atom contacts \cite{Krans},
the tail of the conductance histogram peak at $G_0$ was shown 
to be extended up to 1.2 $G_0$, which may be interpreted 
as an indication of at least two additional eigenchannels~\cite{J=2} 
with a small contribution of $\sim 0.1\, G_0$ for each channel. 

In conductance calculations using finite-sized electrodes, 
the level broadening $E_T$ should satisfy the following constraints;
to retain the metallic nature, 
$E_T$ should be larger than single particle level spacing 
$\delta \epsilon$ ($\sim 0.02$ eV near $E_F$), while
it should be smaller than $\Gamma$ and $\Delta E$ ($\sim 0.3$ eV).
When $E_T$ varies between 0.03 and 0.15 eV, we verify that
$G$ for odd $L$ is almost independent (to within a few percent). 
For even $L$, on the other hand, 
the resonance width $\Gamma$ depends weakly on $E_T$
and thus the choice of $E_T$ affects $G$ quantitatively
although the even-odd behavior remains robust.
We also point out that $\Gamma$ depends on the sharpness 
of the tip structure.
For the pyramid-shape (100) tip, which is less sharper than the (111) tip,
we find that $\Gamma$ increases by about 40\% and the overlap of
resonances is enhanced, leading to weaker even-odd behavior
and less accurate conductance quantization for odd $L$.
Recent calculations \cite{Kobayashi} for an $L=3$ wire with flat
electrodes showed that the energy dependence of the transmission
coefficient is almost negligible, implying an even larger $\Gamma$
of about 1 eV.

Finally, we note that when the exact inversion (or mirror
reflection) symmetry is relaxed, the even-odd behavior
still occurs because the on(off)-resonance for odd (even)
$L$ arises from the sharp tip structure and charge neutrality.
However, the conductance quantization for odd $L$
is weakened, as illustrated clearly in Ref.~\cite{Yeyati}. 
Recent experiments \cite{Rodrigues} have demonstrated that the electrodes
of atomic contacts tend to be aligned with symmetric lattice axes. 
Here we remark that a similar conductance oscillation due to the
resonance states is realized for heterogeneous systems \cite{Lang}, 
where wires and electrodes are made of different atomic elements.
The origin of the resonance is however different;
in heterogeneous systems, the charge transfer between the wires and
electrodes is likely to occur and the resulting Schottky-like potential 
barriers at both the wire ends generate the resonance states
even if the electrodes are flat. 

In summary, a first-principles method is
implemented to calculate conductances through Na wires, with
the aid of the Friedel sum rule.
Unlike conventional QPCs, the transport in monatomic Na wires
has resonant character due to the sharp tip structure.
Combined with the charge neutrality,
the resonance states lead to robust quantized conductances
when the number ($L$) of atoms in the wire is an odd number.
For even $L$, the off-resonance transport is realized with $G$ smaller
than $G_0$.
We suggest that the same even-odd behavior of conductances
may appear in other monovalent atomic wires.

We thank Y.-G. Jin for his advice on the efficient real-space
multigrid method.
We also acknowledge helpful discussions with profs. C. S. Kim and J. Yu.
This work was supported by the QSRC at Dongkuk University.

\begin{references}

\bibitem{Ruitenbeek99cond-mat} J. M. van Ruitenbeek, cond-mat/9910394.
\bibitem{Olesen} L. Olesen {\it et al.}, Phys. Rev. Lett. {\bf 72}, 
2251 (1994).
\bibitem{Wees} B. J. van Wees {\it et al.},
Phys. Rev. Lett. {\bf 60}, 848 (1988);
D. A. Wharam {\it et al.},
J. Phys. C {\bf 21}, L209 (1988).
\bibitem{Glazman} L. I. Glazman {\it et al.},
JETP Lett. {\bf 48}, 238 (1988).
\bibitem{Krans95PRL} J. M. Krans {\it et al.},
Phys. Rev. Lett. {\bf 74}, 2146 (1995). 
\bibitem{Yeyati} A. Levy Yeyati, A. Mart\'{\i}n-Rodero, and F. Flores,
Phys. Rev. B {\bf 56}, 10369 (1997);
J. C. Cuevas {\it et al.}, 
Phys. Rev. Lett. {\bf 81}, 2990 (1998).
\bibitem{Emberly} E. G. Emberly and G. Kirczenow,
Phys. Rev. B {\bf 60}, 6028 (1999).
\bibitem{Brandbyge} M. Brandbyge, N. Kobayashi, and M. Tsukada,
Phys. Rev. B {\bf 60}, 17064 (1999).
\bibitem{Kobayashi} N. Kobayashi, M. Brandbyge, and M. Tsukada,
Surf. Sci. {\bf 433-435}, 854 (1999); 
Phys. Rev. B {\bf 62}, 8430 (2000).
\bibitem{Nakamura} 
Monatomic Na wires with $L>1$ [Fig.~\ref{fig1}(a)]
have not been reported experimentally yet, while 
a possibility of 3 atom-long Na wires has been demonstrated 
theoretically;
A. Nakamura {\it et al.}, Phys. Rev. Lett. {\bf 82}, 1538 (1999).
\bibitem{Datta} S. Datta and W. Tian,
Phys. Rev. B {\bf 55}, R1914 (1997).
\bibitem{Lee} H.-W. Lee,
Phys. Rev. Lett. {\bf 82}, 2358 (1999).
\bibitem{vandenBrom} H. E. van den Brom and J. M. van Ruitenbeek,
Phys. Rev. Lett. {\bf 82}, 1526 (1999).
\bibitem{Jin} Y.-G. Jin, J.-W. Jeong, and K. J. Chang,
Physica B {\bf 273-274}, 1003 (1999).
\bibitem{Troullier_Kleinman} N. Troullier and J. L. Martins,
Phys. Rev. B {\bf 43}, 1993 (1991);
L. Kleinman and D. M. Bylander,
Phys. Rev. Lett. {\bf 48}, 1425 (1982).
\bibitem{Payne} M. C. Payne {\it et al.}, Rev. Mod.
Phys. {\bf 64}, 1045 (1992).
\bibitem{Sablikov} V. A. Sablikov, S. V. Polyakov,
and M. B\"{u}ttiker, Phys. Rev. B {\bf 61}, 13763 (2000).
\bibitem{Harrison}
See for instance W. A. Harrison, {\it Solid State Theory},
Fig.~2.47 (Dover, New York, 1979).
\bibitem{neutrality}
We remark that
the reference system itself has doubly degenerate (one with odd and
the other with even parity) resonance states 
(near black circles in Fig. \ref{fig3}), which are mostly
localized near the apex atoms of the electrodes.
When the two electrodes are connected to make, for example, an $L=5$ wire,
the doubly degenerate resonance states near $E_F-0.3$ eV are mixed
with the wire states to generate three filled and one half-filled resonance
states.
Then, the number of {\it new} filled resonance states is $1\frac{1}{2}$.
\bibitem{Lee01PRB} H.-W. Lee and C. S. Kim, 
Phys. Rev. B {\bf 63}, 075306 (2001);
J. Korean Phys. Soc. {\bf 37}, 137 (2000).
\bibitem{Scheer} E. Scheer {\it et al.},
Nature (London) {\bf 394}, 154 (1998).
\bibitem{Hakkinen} See Ref.~[12] in H. H$\rm\ddot{a}$kkinen, R. N. Barnett, 
and U. Landman, J. Phys. Chem. B {\bf 103}, 8814 (1999).
\bibitem{Yanson} A. I. Yanson {\it et al.},
Nature (London) {\bf 395}, 783 (1998).
\bibitem{Ohnishi} H. Ohnishi, Y. Kondo, and K. Takayanagi,
Nature (London) {\bf 395}, 780 (1998).
\bibitem{gold_wire_comment} 
For rotating zigzag gold wires proposed by D. S\'{a}nchez-Portal {\it et al}.
[Phys. Rev. Lett. {\bf 83}, 3884 (1999)], 
$L$ should be an odd number to be consistent with the large interatomic
spacing observed in Ref.~\protect\cite{Ohnishi}.
In addition, there may be a possibility that gold wires are contaminated
by atoms with low atomic numbers.
\bibitem{Krans} J. M. Krans {\it et al.},
Nature (London) {\bf 375}, 767 (1995).
\bibitem{J=2} 
With only one additional channel, $G$ is still $G_0$ 
due to the identity relation,
$\sin^2[(\pi/2)\alpha]+\sin^2[(\pi/2)(1-\alpha)]=1$.
\bibitem{Rodrigues} V. Rodrigues {\it et al}.,
Phys. Rev. Lett. {\bf 85}, 4124 (2000).
\bibitem{Lang} N. D. Lang, Phys. Rev. Lett. {\bf 79}, 1357 (1997);
N. D. Lang and Ph. Avouris, Phys. Rev. Lett. {\bf 84}, 358 (2000).

\end{references}
\end{document}